\documentclass[aps,prd,twocolumn,superscriptaddress,nofootinbib]{revtex4}

\usepackage{graphicx}
\usepackage{dcolumn}
\usepackage{bm}

\begin{document}

\title{Quark Flavor Mixings from Hierarchical Mass Matrices}

\author{Rohit Verma}

\email{rohitverma@ihep.ac.cn}

\affiliation{Institute of High Energy Physics, Chinese Academy of Sciences, Beijing, 100049, China}
\affiliation{Rayat Institute of Engineering and Information Technology, Ropar 144533, India}

\author{Shun Zhou}

\email{zhoush@ihep.ac.cn}

\affiliation{Institute of High Energy Physics, Chinese Academy of Sciences, Beijing, 100049, China}
\affiliation{Center for High Energy Physics, Peking University, Beijing 100080, China}

\date{\today}

\begin{abstract}
In this paper, we extend the Fritzsch ansatz of quark mass matrices while retaining their hierarchical structures and show that the main features of the Cabibbo-Kobayashi-Maskawa (CKM) matrix $V$, including $|V^{}_{us}| \simeq |V^{}_{cd}|$, $|V^{}_{cb}| \simeq |V^{}_{ts}|$ and $|V^{}_{ub}|/|V^{}_{cb}| < |V^{}_{td}|/|V^{}_{ts}|$, can be well understood. This agreement is observed especially when the mass matrices have non-vanishing $(1,3)$ and $(3,1)$ off-diagonal elements. The phenomenological consequences of these for the allowed texture content and gross structural features of `hierarchical' quark mass matrices are addressed from a model independent prospective under the  assumption of factorizable phases in these. The approximate and analytical expressions of the CKM matrix elements are derived, and a detailed  analysis reveals that such  structures are in good agreement with the observed quark flavor mixing angles and the CP-violating phase at the $1\sigma$ level and call upon a further investigation of the realization of these structures from a top-down prospective.
\end{abstract}


\maketitle

\section{Introduction}

The ATLAS~\cite{Aad:2012tfa} and CMS~\cite{Chatrchyan:2012xdj} experiments at the CERN Large Hadron Collider (LHC) have recently discovered a scalar particle of mass $m^{}_h \approx 125~{\rm GeV}$, which is likely to be the Higgs boson in the standard model (SM), witnessing a great success of the SM in describing the electroweak interactions. However, the flavor structures of Yukawa interactions between the Higgs boson and the fermions are poorly understood. For instance, the strong hierarchy among quark masses, the smallness of quark flavor mixing angles and the origin of CP violation remain to be explained. As the quark masses and flavor mixing parameters have been precisely measured, it is an important and urgent task to explore the flavor dynamics of quarks~\cite{Fritzsch:1999ee,Gupta:2013yha,Xing:2014sja}.

To be explicit, we evaluate the running quark masses at the energy scale of $M^{}_Z = 91.2~{\rm GeV}$. The mass ratios are found to be~\cite{Xing:2011aa}
\begin{eqnarray}\label{eq:massratio}
&& \frac{m^{}_u}{m^{}_c} = 2.2^{+1.1}_{-0.8} \times 10^{-3} \; , ~ \frac{m^{}_c}{m^{}_t} = 3.7^{+0.3}_{-0.5} \times 10^{-3} \; , \nonumber \\
&& \frac{m^{}_d}{m^{}_s} = 4.9^{+2.4}_{-1.8}  \times 10^{-2} \; , ~
\frac{m^{}_s}{m^{}_b} = 2.0^{+0.7}_{-0.5}  \times 10^{-2} \; , ~~
\end{eqnarray}
and the absolute masses of the third-family quarks are $m^{}_b = 2.86^{+0.16}_{-0.06}~{\rm GeV}$ and $m^{}_t = 172.1^{+1.2}_{-1.2}~{\rm GeV}$. The simple relations $m^{}_u/m^{}_c \approx m^{}_c/m^{}_t \approx \lambda^4$ and $m^{}_d/m^{}_s \approx m^{}_s/m^{}_b \approx \lambda^2$ are valid within the uncertainties, where $\lambda \equiv \sin \theta^{}_{\rm C} \approx 0.225$ with $\theta^{}_{\rm C}$ being the well-known Cabibbo angle. Therefore, one can observe that quark mass spectra exhibit a strong hierarchy, i.e., $m^{}_u \ll m^{}_c \ll m^{}_t$ and $m^{}_d \ll m^{}_s \ll m^{}_b$. On the other hand, the quark flavor mixing is described by the CKM matrix $V$~\cite{Cabibbo:1963yz, Kobayashi:1973fv}, and the moduli of its nine matrix elements $|V^{}_{ij}|$ for $i = u, c, t$ and $j = d, s, b$ have been determined from current experimental data with high precisions~\cite{Agashe:2014kda}. A strongly hierarchical structure of CKM matrix elements is also observed, namely, $|V^{}_{tb}| > |V^{}_{ud}| > |V^{}_{cs}| \gg |V^{}_{us}| > |V^{}_{cd}| \gg |V^{}_{cb}| > |V^{}_{ts}|
\gg |V^{}_{td}| > |V^{}_{ub}|$.
These strong hierarchies in quark mass spectra and flavor mixings are probably correlated with each other.

Unfortunately, we have not yet had a convincing flavor theory to realize such a correlation. Therefore, phenomenological approaches on specific textures of quark mass matrices have been widely adopted to obtain vital clues on the underlying flavor dynamics \cite{Froggatt:1978nt, Dimopoulos:1991yz, Chiu:2000gw, Grimus:2004hf,Ma:2004pt, Lavoura:2007dw, EmmanuelCosta:2011jq, Feruglio:2015jfa}. However, experience shows that there exist a vast number of texture zero structures, \cite{Ahuja:2009jj, Harigaya:2012bw, Verma:2013cza, Fakay:2013gf, Verma:2014woa, Verma:2014lpa, Ludl:2015lta, Cebola:2015dwa, Liao:2015hya, Kim:2015zla, Gautam:2015kya, Ahuja:2015crb, Cebola:2016jhz, Tanimoto:2016rqy} both for the quark and lepton sectors, which in principle, may not be ruled out by experiments. This suggests that the texture approach alone is perhaps not sufficient enough to account for the flavor dynamics and one may require to use additional assumptions (based on observations) to extract some useful  information on the gross possible structures of these matrices from a model-independent prospective, particularly, for the quark sector.

A natural explanation for quark flavor dynamics requires a translation of the observed hierarchies in the quark masses and mixing angles onto the corresponding matrix elements in these mass matrices.    One typical example is the Fritzsch mass matrices with six texture zeros~\cite{Fritzsch:1977vd, TNYA:TNYA2958}
\begin{eqnarray}\label{eq:fritzsch}
M^{}_{\rm q} = \left(\matrix{{\bf 0} & A^{}_{\rm q} & {\bf 0} \cr A^*_{\rm q} & {\bf 0} & B^{}_{\rm q} \cr {\bf 0} & B^*_{\rm q} & C^{}_{\rm q} \cr } \right) \; ,
\end{eqnarray}
where ${\rm q} = {\rm u}$ or ${\rm d}$ denotes the up-type or down-type quark mass matrix, and a pair of off-diagonal texture zeros in the Hermitian or symmetric matrix are usually counted as one zero. As a consequence of texture zeros \cite{TNYA:TNYA2958, Fritzsch:1977vd, Fritzsch:1979zq, Fritzsch:1999rb, Fritzsch200363, Gupta:2009ur, Mahajan:2009wd, Verma:2009gf, Verma:2010jy,  Ponce:2013nsa, Xing:2015sva}, a few simple but instructive relations among flavor mixing angles and quark mass ratios can be established \cite{Gatto:1968ss, Fritzsch:1979zq, Ramond:1993kv, Hall1993164,  Barbieri199993,  Roberts2001358, Fritzsch200363, Kim:2004ki, Branco:2006wv} and tested in experiments. Unfortunately, the Fritzsch ansatz failed to reconcile a large top-quark mass with the observed value of $|V^{}_{cb}|$ \cite{Harari1987586, Albright1989171}.

Possible extensions of the Fritzsch ansatz have been considered in the literature~\cite{Fritzsch:1999ee}. For instance, a nonzero element in the $(2,2)$ position of $M^{}_{\rm q}$, i.e.,
\begin{equation}\label{eq:fritzschlike}
M^{}_{\rm q} = \left(\matrix{{\bf 0} & A^{}_{\rm q} & {\bf 0} \cr A^*_{\rm q} & D^{}_{\rm q} & B^{}_{\rm q} \cr {\bf 0} & B^*_{\rm q} & C^{}_{\rm q} \cr } \right) \; ,
\end{equation}
has been proved to be important to accommodate the experimental data on quark flavor mixings~\cite{Du:1992iy}. A remarkable difference between the Fritzsch ansatz in Eq.~(\ref{eq:fritzsch}) and the parallel four-zero texture in Eq.~(\ref{eq:fritzschlike}) is that a hierarchical structure is not allowed in the latter case~\cite{Fritzsch:2002ga, Xing:2003yj, Verma:2009gf, Verma:2010jy}. Even more, it has recently been found in Ref.~\cite{Xing:2015sva} that the relation $|D^{}_{\rm q}| \simeq |B^{}_{\rm q}| \simeq |C^{}_{\rm q}|$ is valid. By contrast, the Fritzsch ansatz with $|A^{}_{\rm q}| \ll |B^{}_{\rm q}| \ll |C^{}_{\rm q}|$ seems to reflect the strong hierarchies of quark mass spectra and flavor mixings in a more transparent and natural way. However, one has not been able to find \cite{Sharma:2015gfa} any strongly hierarchical structure for quark mass matrices which explains the observed flavor mixings.

The purpose of this paper is to extend the Fritzsch mass matrices of Eq.~(\ref{eq:fritzsch}) assuming hierarchical structures for the corresponding mass matrices following a natural mixing among the three quark generations, see Eq.(\ref{eq:hierarchical}). However, a texture zero at the diagonal $(3, 3)$ position disturbs such a natural hierarchy among the mass matrix elements and is therefore ruled out in this context. Particularly, we find that the non-vanishing matrix elements at the off-diagonal $(1, 3)$ and $(3, 1)$ positions are crucial in retaining a natural mass matrix hierarchy. For example, one may consider nonzero elements in both $(1, 3)$ and $(2, 2)$ positions of $M^{}_{\rm u}$, while $M^{}_{\rm d}$ takes the form of the Fritzsch ansatz:
\begin{eqnarray}\label{eq:caseI}
M^{}_{\rm u} &=& \left(\matrix{{\bf 0} & A^{}_{\rm u} & F^{}_{\rm u} \cr A^*_{\rm u} & D^{}_{\rm u} & B^{}_{\rm u} \cr F^*_{\rm u} & B^*_{\rm u} & C^{}_{\rm u} \cr } \right) \; , \nonumber \\
M^{}_{\rm d} &=& \left(\matrix{{\bf 0} & A^{}_{\rm d} & {\bf 0} \cr A^*_{\rm d} & {\bf 0} & B^{}_{\rm d} \cr {\bf 0} & B^*_{\rm d} & C^{}_{\rm d} \cr } \right) \; .
\end{eqnarray}
or vice versa. Another possibility can be obtained from the first type by just moving the nonzero element in the $(1, 3)$ position of $M^{}_{\rm u}$ to the same place of $M^{}_{\rm d}$, i.e.,
\begin{eqnarray}\label{eq:caseII}
M^{}_{\rm u} &=& \left(\matrix{{\bf 0} & A^{}_{\rm u} & {\bf 0} \cr A^*_{\rm u} & D^{}_{\rm u} & B^{}_{\rm u} \cr {\bf 0} & B^*_{\rm u} & C^{}_{\rm u} \cr } \right) \; , \nonumber \\
M^{}_{\rm d} &=& \left(\matrix{{\bf 0} & A^{}_{\rm d} &  F^{}_{\rm d} \cr A^*_{\rm d} & {\bf 0} & B^{}_{\rm d} \cr F^*_{\rm d} & B^*_{\rm d} & C^{}_{\rm d} \cr } \right) \; .
\end{eqnarray}
or vice versa. More examples of such such viable texture structures were recently \cite{Ludl:2015lta} obtained for symmetric mass matrices. However, the study was mainly limited to a numerical investigation only while a desirable phenomenological explanation of these was not explored and requires an urgent investigation. We undertake a phenomenological study of such texture structures, when $M_{\rm q}(3,3)\ne0$, and find that these have a common unique feature of displaying strong hierarchy among the mass matrix elements and that these features can still be retained for Hermitian mass matrices within the framework of the Standard Model.

It is worth stressing that the matrix elements $D^{}_{\rm q}$ of $M^{}_{\rm q}$ can not be vanishing simultaneously, for texture specific hierarchical mass matrices, due to the constraint from the precise measurement of $|V^{}_{cb}|$, as we shall show later. In addition, the nonzero element $F^{}_{\rm u}$ or $F^{}_{\rm d}$ is indispensable to fit the experimental value of $|V^{}_{ub}|$ while preserving the hierarchical structures of quark mass matrices. In Sec. II, based on the weak-basis transformation, we argue that the non-parallel four-zero textures in Eqs.~(\ref{eq:caseI}) and (\ref{eq:caseII}) are as general as the parallel four-zero texture in Eq.~(\ref{eq:fritzschlike}). Sec. III is devoted to an approximate and analytical approach to diagonalize hierarchical quark mass matrices. The expressions of CKM matrix elements have been derived. In Sec. IV, we explore the allowed regions of model parameters by performing a complete numerical analysis and find an excellent agreement with the experimental data. Finally, we summarize in Sec. V.

\section{Weak Basis transformations}

In the SM, the Yukawa interactions between the Higgs boson and the fermions play a very important role. First, for a light Higgs boson of mass $m^{}_h \approx 125~{\rm GeV}$, the dominant decay channel is $h \to b \overline{b}$ and the decay width is governed by the Yukawa coupling of bottom quark. Second, the quark mass spectra, flavor mixing angles and CP violation are determined by the complex Yukawa coupling matrices. As the right-handed quark fields are gauge singlets, it is always possible to make the quark mass matrices $M^{}_{\rm u}$ and $M^{}_{\rm d}$ Hermitian through a unitary transformation of right-handed quark fields in the flavor space, if there is no flavor-changing right-handed current as in the SM. The diagonalizations of $M^{}_{\rm u}$ and $M^{}_{\rm d}$ lead to quark mass eigenvalues $V^\dagger_{\rm u} M^{}_{\rm u} V^{}_{\rm u} = {\rm diag}\{m^{}_u, m^{}_c, m^{}_t\}$ and $V^\dagger_{\rm d} M^{}_{\rm d} V^{}_{\rm d} = {\rm diag}\{m^{}_d, m^{}_s, m^{}_b\}$. Moreover, the mismatch between those diagonalizations results in the CKM matrix $V = V^\dagger_{\rm u} V^{}_{\rm d}$, which enters into the flavor-changing charged-current interactions and thus can be precisely measured in experiments.

It is evident that a weak-basis transformation, defined as $M^{}_{\rm u} \to W M^{}_{\rm u} W^\dagger$ and $M^{}_{\rm d} \to W M^{}_{\rm d} W^\dagger$ where $W$ is a $3\times 3$ unitary matrix, keeps the CKM matrix $V$ unchanged. However, it indeed changes the flavor structure of quark mass matrices. Through the weak-basis transformations, one can prove that the quark mass matrices can be recast into the following forms with three texture zeros:
\begin{eqnarray}\label{eq:Branco}
M^{}_{\rm u} = \left(\matrix{{\bf 0} & \times & {\bf 0} \cr \times & \times & \times \cr {\bf 0} & \times & \times \cr } \right) \; , ~
M^{}_{\rm d} = \left(\matrix{{\bf 0} & \times & \times \cr \times & \times & \times \cr \times & \times & \times \cr } \right) \; , ~~~
\end{eqnarray}
as demonstrated in Ref.~\cite{Branco:1988iq, Branco:1999nb} or
\begin{eqnarray}\label{eq:FX}
M^{}_{\rm u} = \left(\matrix{{\bf 0} & \times & {\bf 0} \cr \times & \times & \times \cr {\bf 0} & \times & \times \cr } \right) \; , ~
M^{}_{\rm d} = \left(\matrix{\times & \times & {\bf 0} \cr \times & \times & \times \cr {\bf 0} & \times & \times \cr } \right) \; , ~~~
\end{eqnarray}
as pointed out in Ref.~\cite{Fritzsch:1997fw}. Notice that Eqs.~(\ref{eq:Branco}) and (\ref{eq:FX}) are valid as well when the texture structures of $M^{}_{\rm u}$ and $M^{}_{\rm d}$ are exchanged. Note also that the crosses in the mass matrices in Eqs.~(\ref{eq:Branco}) and (\ref{eq:FX}) stand for nonzero elements, and the Hermiticity is implied. The texture zeros stemming from weak-basis transformations carry no physical information, but additional ones lead to correlations among quark masses and flavor mixing angles. Two comments are in order:
\begin{itemize}
\item In Eq.~(\ref{eq:FX}), the element in the $(1,1)$ position of $M^{}_{\rm d}$ has been found to be redundant~\cite{Verma:2013qta, Sharma:2015soa, Sharma:2014tea} in the sense that the quark mass matrices are compatible with experimental data even in the limit of $(M^{}_{\rm d})^{}_{11} = 0$. In this limit, we obtain the Fritzsch-like mass matrices with four texture zeros in Eq.~(\ref{eq:fritzschlike}). Therefore, the parallel four-zero textures can be derived from the most general ones in Eq.~(\ref{eq:FX}) under the minimal additional assumption of a vanishing $(1,1)$ diagonal element. Furthermore, most of the recent investigations involving texture specific quark mass matrices have been limited to parallel texture zeros \cite{Sharma:2014tea, Sharma:2015gfa,Sharma:2015soa} or based on numerical investigations only \cite{Ludl:2015lta}.

\item A comparison of mass matrices in  Eqs.~(\ref{eq:caseI}), (\ref{eq:caseII}) and Eq.~(\ref{eq:Branco}) reveals that one matrix element in the diagonal position of $M^{}_{\rm u}$ or $M^{}_{\rm d}$ in Eq.~(\ref{eq:Branco}) has been taken to be zero. Hence, the non-parallel four-zero textures are as general as the parallel ones in Eq.~(\ref{eq:fritzschlike}) and deserve a dedicated study.
\end{itemize}

In the following section, we validate an approximate and analytical diagonalization of such non-parallel quark mass matrices and consequently establish simple relations between the CKM matrix elements and quark masses. With the help of these relations, we can translate the observed strong hierarchies among the quark masses and mixing angles into the hierarchical structures of quark mass matrices. Such a translation is not possible for the parallel four-zero textures in Eq.~(\ref{eq:fritzschlike}) where $|V_{ub}|$ is obtained from a next-to-leading order mixing of the neighboring quark generations owing to a suppression of a leading-order mixing among the first and third quark generations through the position of texture zeros in these.

\section{Analytical Results}

In view of the strong hierarchy of quark masses, we expect naturally the hierarchical structures of quark mass matrices, which in the most general case can be formulated as follows
\begin{eqnarray}\label{eq:hierarchical}
M^{}_{\rm q} &=& \left(\matrix{ e^{}_{\rm q} & a^{}_{\rm q} e^{{\rm i} \alpha^{}_{\rm q}} & f^{}_{\rm q} e^{{\rm i} \gamma^{}_{\rm q}} \cr a^{}_{\rm q} e^{-{\rm i} \alpha^{}_{\rm q}} & d^{}_{\rm q} & b^{}_{\rm q} e^{{\rm i} \beta^{}_{\rm q}} \cr f^{}_{\rm q} e^{-{\rm i} \gamma^{}_{\rm q}} & b^{}_{\rm q} e^{-{\rm i} \beta^{}_{\rm q}}  & c^{}_{\rm q} \cr } \right) \nonumber \\
&\sim& \left(\matrix{ m^{}_1 & \sqrt{m^{}_1 m^{}_2} e^{{\rm i} \alpha^{}_{\rm q}} &\sqrt{m^{}_1 m^{}_3} e^{{\rm i} \gamma^{}_{\rm q}} \cr \sqrt{m^{}_1 m^{}_2} e^{-{\rm i} \alpha^{}_{\rm q}} & m^{}_2 & \sqrt{m^{}_2 m^{}_3} e^{{\rm i} \beta^{}_{\rm q}} \cr \sqrt{m^{}_1 m^{}_3} e^{-{\rm i} \gamma^{}_{\rm q}} & \sqrt{m^{}_2 m^{}_3} e^{-{\rm i} \beta^{}_{\rm q}}  & m^{}_3 \cr } \right) \; , \nonumber \\
\end{eqnarray}
where $m^{}_i$ denotes the mass of $i$-th generation up-type (or down-type) quark for ${\rm q} = {\rm u}$ (or ${\rm d}$), and each matrix element is expressed in terms of its modulus and argument. In this formulation, the hierarchical structure of a mass matrix should be understood as the rightmost one in Eq.~(\ref{eq:hierarchical}) up to a factor of ${\cal O}(1)$ for each nonzero matrix element. The hierarchical structures have been extensively discussed in the literature~\cite{Fritzsch:1999ee,Gupta:2013yha,Xing:2014sja}, for example, we have $|C^{}_{\rm q}| \sim m^{}_3 \gg |B^{}_{\rm q}| \sim \sqrt{m^{}_2 m^{}_3} \gg |A^{}_{\rm q}| \sim \sqrt{m^{}_1 m^{}_2}$ for the original Fritzsch ansatz.

In order to achieve an analytical diagonalization of the mass matrices and realize simple empirical relations among the quark mixing angles and the corresponding mass ratios, we assume that the phases are factorizable in $M^{}_{\rm q}$ i.e. $\gamma^{}_{\rm q} = \alpha^{}_{\rm q} + \beta^{}_{\rm q}$. This also reduces the number of free parameters in the associated quark mass matrices, to be compared with ten physical observables. As a result, one can rewrite the mass matrix as $M_{\rm q} = P^{}_{\rm q} M^{\prime}_{\rm q} P^\dagger_{\rm q}$, where $P^{}_{\rm q} \equiv {\rm diag}\{e^{{\rm i}\alpha^{}_{\rm q}}, 1, e^{-{\rm i} \beta^{}_{\rm q}}\}$ and $M^{\prime}_{\rm q}$ is a real and symmetric matrix. The diagonalization of $M^{\prime}_{\rm q}$ can now be achieved by an orthogonal matrix $O^{}_{\rm q}$, which can be parametrized in terms of three rotation angles $\{\theta^{\rm q}_{12}, \theta^{\rm q}_{13}, \theta^{\rm q}_{23}\}$. Adopting a convenient parametrization $O^{}_{\rm q} \equiv R^{}_{13}(\theta^{\rm q}_{13}) R^{}_{12}(\theta^{\rm q}_{12}) R^{}_{23}(\theta^{\rm q}_{23})$, where $R^{}_{ij}(\theta^{\rm q}_{ij})$ is the rotation matrix in the $ij$-plane with an angle $\theta^{\rm q}_{ij}$ (for $ij = 12$, $13$, $23$), we have $O^{\rm T}_{\rm q} M^{\prime}_{\rm q} O^{}_{\rm q} = \tilde{M}_{\rm q}$, where $\tilde{M}_{\rm q} = {\rm diag}\{ k_{q\rm } m^{}_1, - k_{\rm q} m^{}_2, m^{}_3\}$ is a diagonal matrix with quark mass eigenvalues and $k_{\rm q}=\pm 1$. The diagonalization of the mass matrices using $O^{\rm T}_{\rm q} M^{\prime}_{\rm q} O^{}_{\rm q} = \tilde{M}_{\rm q}$ for the above hierarchical mass matrices, with factorizable phases, requires one negative mass eigenvalue to ensure real values for trigonometric sines of the associated rotation angles $\theta^{\rm q}_{ij}$ obtained in the following discussion.

Due to the hierarchical structure of $M^{}_{\rm q}$, we expect that the rotation angles $\theta^{\rm q}_{ij}$ are associated with the quark mass ratios and thus in general very small. To the leading order of small rotation angles, the orthogonal matrix is given by
\begin{eqnarray}\label{eq:Oq}
O^{}_{\rm q} \approx \left(\matrix{c^{\rm q}_{12} & s^{\rm q}_{12} & s^{\rm q}_{13} \cr -s^{\rm q}_{12} & c^{\rm q}_{12} c^{\rm q}_{23} & s^{\rm q}_{23} \cr -s^{\rm q}_{13} + s^{\rm q}_{12} s^{\rm q}_{23} & -s^{\rm q}_{23} & c^{\rm q}_{23}}\right) \; ,
\end{eqnarray}
where $s^{\rm q}_{ij} \equiv \sin \theta^{\rm q}_{ij}$ and $c^{\rm q}_{ij} \equiv \cos \theta^{\rm q}_{ij}$ have been defined. Then, the CKM matrix is determined by
\begin{equation}\label{eq:CKM}
V = O^{\rm T}_{\rm u} P^\dagger_{\rm u} P_{\rm d} O^{}_{\rm d} = O^{\rm T}_{\rm u} \left(\matrix{e^{-{\rm i}\phi^{}_1} & 0 & 0 \cr 0 & 1 & 0 \cr 0 & 0 & e^{{\rm i}\phi^{}_2}}\right) O^{}_{\rm d} \; , ~~
\end{equation}
where $\phi^{}_1 \equiv \alpha^{}_{\rm u} - \alpha^{}_{\rm d}$ and $\phi^{}_2 \equiv \beta^{}_{\rm u} - \beta^{}_{\rm d}$.

The final task is to calculate the rotation angles $\theta^{\rm q}_{ij}$ from the quark mass matrices. Using $O^{\rm T}_{\rm q} M^{\prime}_{\rm q} O^{}_{\rm q} = \tilde{M}_{\rm q}$, it is straightforward to verify
\begin{eqnarray}\label{eq:abc}
c^{}_{\rm q} &\approx& m^{}_3 \; , \nonumber \\
b^{}_{\rm q} &\approx& m^{}_3 s^{\rm q}_{23} \; , \nonumber \\
a^{}_{\rm q} &\approx& - k_{\rm q} m^{}_2 s^{\rm q}_{12} + m^{}_3 s^{\rm q}_{13} s^{\rm q}_{23} \; ,
\end{eqnarray}
and
\begin{eqnarray}\label{eq:edf}
f^{}_{\rm q} &\approx& m^{}_3 s^{\rm q}_{13}\; , \nonumber \\
d^{}_{\rm q} &\approx& - k_{\rm q} m^{}_2 + m^{}_3 (s^{\rm q}_{23})^2 \; , \nonumber \\
e^{}_{\rm q} &\approx&  k_{\rm q} m^{}_1 - k_{\rm q} m^{}_2 (s^{\rm q}_{12})^2 \; ,
\end{eqnarray}
in the leading-order approximation. Taking $e^{}_{\rm q} = 0$, as for the most general mass matrices in Eq.~(\ref{eq:Branco}), one obtains
\begin{eqnarray}\label{eq:rotation}
s^{\rm q}_{12} &\approx&  \sqrt{\frac{m^{}_1}{m^{}_2}}\; , \nonumber \\
s^{\rm q}_{13} &\approx& \varepsilon^{}_{\rm q} \sqrt{\frac{m^{}_1}{m^{}_3}} \; , \nonumber \\
s^{\rm q}_{23} &\approx&  \sqrt{\frac{d^{}_{\rm q} + k_{\rm q} m^{}_2}{m^{}_3}} \; ,
\end{eqnarray}
where we have introduced $\varepsilon^{}_{\rm q} \equiv f^{}_{\rm q}/\sqrt{m^{}_1 m^{}_{3}}\sim {\cal O}(1)$ in agreement with hierarchical structures in Eq.~({\ref{eq:hierarchical}) wherein $f_{\rm q}\sim {\cal O}\sqrt{m_1 m_3}$. Note that the rotation angles have been expressed in terms of quark masses and four free parameters $\{\varepsilon^{}_{\rm u}, d^{}_{\rm u}\}$ and $\{\varepsilon^{}_{\rm d}, d^{}_{\rm d}\}$. They can be substituted into Eq.~(\ref{eq:abc}) to obtain the other matrix elements of $\tilde{M}_{\rm q}$. According to Eqs.~(\ref{eq:Oq}) and (\ref{eq:CKM}), it is easy to derive the following simple relations for the off-diagonal elements of the CKM matrix:
\begin{eqnarray}\label{eq:approx}
V^{}_{us} &\approx& \sqrt{\frac{m^{}_d}{m^{}_s}} e^{ - {\rm i} \phi^{}_1} - \sqrt{\frac{m^{}_u}{m^{}_c}} \; , \nonumber \\
V^{}_{cd} &\approx& \sqrt{\frac{m^{}_u}{m^{}_c}} e^{ - {\rm i} \phi^{}_1} - \sqrt{\frac{m^{}_d}{m^{}_s}} \; , \nonumber \\
V^{}_{cb} &\approx& \sqrt{\frac{d^{}_{\rm d} + k_{\rm d} m^{}_s}{m^{}_b}} - \sqrt{\frac{d^{}_{\rm u} + k_{\rm u} m^{}_c}{m^{}_t}} e^{{\rm i}\phi^{}_2} \; , \nonumber \\
V^{}_{ts} &\approx& \sqrt{\frac{d^{}_{\rm u} + k_{\rm u} m^{}_c}{m^{}_t}} - \sqrt{\frac{d^{}_{\rm d} + k_{\rm d} m^{}_s}{m^{}_b}} e^{{\rm i}\phi^{}_2} \; , \nonumber \\
V^{}_{ub} &\approx& + \varepsilon^{}_{\rm d} \sqrt{\frac{m^{}_d}{m^{}_b}} e^{ - {\rm i} \phi^{}_1} - \varepsilon^{}_{\rm u} \sqrt{\frac{m^{}_u}{m^{}_t}} e^{{\rm i} \phi^{}_2} - \sqrt{\frac{m^{}_u}{m^{}_c}} V^{}_{cb} \; , \nonumber \\
V^{}_{td} &\approx& + \varepsilon^{}_{\rm u} \sqrt{\frac{m^{}_u}{m^{}_t}} e^{ - {\rm i} \phi^{}_1} - \varepsilon^{}_{\rm d} \sqrt{\frac{m^{}_d}{m^{}_b}} e^{{\rm i} \phi^{}_2} - \sqrt{\frac{m^{}_d}{m^{}_s}} V^{}_{ts} \; . \nonumber \\
\end{eqnarray}
Some remarks on the above approximate and analytical results are helpful. First of all, the relation $|V^{}_{us}| \approx |V^{}_{cd}| \approx \sqrt{m^{}_d/m^{}_s} \approx \lambda$ holds as in the case of Fritzsch mass matrices. Second, if $d^{}_{\rm u} = 0$ is assumed, we can derive a lower bound on $|V^{}_{cb}|$ from Eq.~(\ref{eq:approx}), namely,
\begin{eqnarray}\label{eq:Vcb}
|V^{}_{cb}| \gtrsim  \sqrt{\frac{m^{}_s}{m^{}_b}} - \sqrt{\frac{m^{}_c}{m^{}_t}} \gtrsim 0.06 \; ,
\end{eqnarray}
where the quark mass ratios in Eq.~(\ref{eq:massratio}) have been used. The lower bound in Eq.~(\ref{eq:Vcb}) is in contradiction with the experimental value $|V^{}_{cb}| = 0.0414 \pm 0.0012$ that has been precisely measured in the exclusive and inclusive semileptonic decays of $B$ mesons into charm~\cite{Agashe:2014kda}. Hence, it is evident why the Fritzsch ansatz with a hierarchical structure and $d^{}_{\rm u} =d^{}_{\rm d}= 0$ is excluded. A nonzero $d^{}_{\rm q}$ is of crucial importance to accommodate the experimental data. Third, we observe from the last two lines in Eq.~(\ref{eq:approx}) that
\begin{eqnarray}\label{eq:Vub}
\frac{|V^{}_{ub}|}{|V^{}_{cb}|} &\approx& \sqrt{\frac{m^{}_u}{m^{}_c}} = 0.0469^{+0.0105}_{-0.0095}\; , ~~~ \nonumber \\
\frac{|V^{}_{td}|}{|V^{}_{ts}|} &\approx& \sqrt{\frac{m^{}_d}{m^{}_s}} = 0.221^{+0.049}_{-0.045}\; , ~~~
\end{eqnarray}
where the quark mass ratios are used and $\varepsilon^{}_{\rm u} = \varepsilon^{}_{\rm d} = 0$ is assumed. The second relation in Eq.~(\ref{eq:Vub}) is compatible with $|V^{}_{td}|/|V^{}_{ts}| = 0.216 \pm 0.011$, which has been extracted from the mass difference for the neutral $B$ mesons in the $B^0_d$-$\overline{B}^0_d$ system and that in the $B^0_s$-$\overline{B}^0_s$ system~\cite{Agashe:2014kda}. But, the estimated value of $|V^{}_{ub}|/|V^{}_{cb}| \approx 0.0469$ in Eq.~(\ref{eq:Vub}) is smaller than the observed one $0.0857$~\cite{Agashe:2014kda} almost by a factor of two.

Based on the above observations, we conclude that $\varepsilon^{}_{\rm u}$ and $\varepsilon^{}_{\rm d}$ cannot be vanishing simultaneously, and likewise for $d^{}_{\rm u}$ and $d^{}_{\rm d}$. To emphasize this further, we carry out a detailed investigation of mass matrix structures in Eqs.~(\ref{eq:caseI}) and (\ref{eq:caseII}) as the minimal extension of the Fritzsch ansatz where hierarchical structure is retained. The analytical results in Eq.~(\ref{eq:approx}) not only hold for the general quark mass matrices in Eqs.~(\ref{eq:Branco}),  (\ref{eq:hierarchical}) but are also applicable to all viable Hermitian texture structures obtained in Ref. \cite{Ludl:2015lta} with $M_{\rm q}(3,3)\ne 0$ under the assumption of factorizable phases in these.

Furthermore, from the expression for $s^{\rm q}_{23}$ in Eq.~(\ref{eq:rotation}), it is evident that $d_{\rm q}=0$ is not allowed for $k_{\rm q}=-1$ corresponding to the negative $m_1$ eigenvalue. As a result, one is compelled to consider $m_2$ negative for the corresponding mass matrices with $d_{\rm d}=0$ in Eq.~(\ref{eq:caseI}) and $d_{\rm u}=0$ in Eq.~(\ref{eq:caseII}). However, a negative sign associated with either of the two mass eigenvalues is consistent with the data for mass matrices with $d_{\rm q}\ne 0$ and it is trivial to check that the allowed parameter space for $d_{\rm q}$ is not very different in such cases. In particular, the modular $d_{\rm q}$ values are of the same order but marginally larger for $k_{\rm q}=-1$ as compared to the $k_{\rm q}=+1$ which is also clearly evident through the expression for $V_{cb}\approx s^{\rm d}_{23}-s^{\rm u}_{23}e^{i\phi_{2}}$ when $d_{\rm d}=0$ (or $d_{\rm u}=0$). As a result, we restrict our analysis with $k_{\rm q}=+1$ for obtaining greater agreement with the notion of `hierarchical' mass matrices assumed in Eq.~(\ref{eq:hierarchical}).

\section{Numerical Analysis}

Now we proceed to explore the parameter space of the non-parallel four-zero textures of quark mass matrices in Eqs.~(\ref{eq:caseI}) and (\ref{eq:caseII}). We use the running quark masses at $M^{}_Z = 91.2~{\rm GeV}$ as inputs
\begin{eqnarray}\label{eq:quark masses}
&~& m^{}_u = 1.38^{+0.42}_{-0.41}~{\rm MeV} \; ,~~~~
m^{}_d = 2.82^{+0.48}_{-0.48}~{\rm MeV}\; , \nonumber \\
&~& m^{}_c = 0.638^{+0.043}_{-0.084}~{\rm GeV} \; , ~~
m^{}_s = 57^{+18}_{-12}~{\rm MeV}\;,\nonumber \\
&~& m^{}_t = 172.1^{+1.2}_{-1.2}~{\rm GeV} \; , ~~~~ \;
m^{}_b = 2.86^{+0.16}_{-0.06}~{\rm GeV}\; , ~~~~
\end{eqnarray}
where the central values with $1\sigma$ errors are taken from Ref.~\cite{Xing:2011aa}. The mass matrices are claimed to be consistent with experimental data if the absolute values of three CKM matrix elements and one inner angle of the unitarity triangle can be reproduced~\cite{Agashe:2014kda}, i.e.,
\begin{eqnarray}\label{eq:constraint}
|V^{}_{us}| &=& 0.22536 \pm 0.00061 \; , \nonumber \\
|V^{}_{ub}| &=& 0.00355 \pm 0.00015 \; , \nonumber \\
|V^{}_{cb}| &=& 0.0414 \pm 0.0012 \; , \nonumber \\
\sin 2\beta &=& 0.682 \pm 0.019 \; ,
\end{eqnarray}
where the $1\sigma$ errors are attached to the best-fit values. At the same time, the allowed regions of free model parameters can be obtained. As usual, three inner angles of the unitarity triangle correspond to the orthogonal condition $V^{}_{ud} V^*_{ub} + V^{}_{cd} V^*_{cb} + V^{}_{td} V^*_{tb} = 0$, e.g.
\begin{eqnarray}
\alpha  &\equiv& \arg \left(-\frac{V^{}_{td} V^*_{tb}}{V^{}_{ud} V^*_{ub}}\right) \; , \nonumber \\
\beta  &\equiv& \arg \left(-\frac{V^{}_{cd} V^*_{cb}}{V^{}_{td} V^*_{tb}}\right) \; , \nonumber \\
\gamma  &\equiv& \arg \left(-\frac{V^{}_{ud} V^*_{ub}}{V^{}_{cd} V^*_{cb}}\right) \; ,
\end{eqnarray}
etc., which can be determined independently from the measurements of CP violation in $B$ decays~\cite{Agashe:2014kda}.

\subsection{Case I}

We consider the quark mass matrices in Eq.~(\ref{eq:caseI}). In this case, we have $f^{}_{\rm d} \propto \varepsilon^{}_{\rm d} = 0$ and $d^{}_{\rm d} = 0$, but $f^{}_{\rm u} \propto \varepsilon^{}_{\rm u} \neq 0$ along with $d^{}_{\rm u} \ne 0$. Hence, the free model parameters include $\{d^{}_{\rm u}, \varepsilon^{}_{\rm u}\}$ and two phases $\{\phi^{}_1, \phi^{}_2\}$. In our numerical calculations, we exactly diagonalize the quark mass matrices, and freely vary the physical ranges of free model parameters $\{d^{}_{\rm u}, \varepsilon^{}_{\rm u}\}$ in their allowed parameter space consistent with Eqs.(\ref{eq:hierarchical}) and (\ref{eq:approx}). Both $\phi^{}_1$ and $\phi^{}_2$ are allowed to freely vary in $[0^\circ, 360^\circ]$ interval. Given quark masses and the values of four model parameters, we are able to calculate the CKM matrix elements. Only when the experimental constraints in Eq.~(\ref{eq:constraint}) are satisfied, we output the values of model parameters and reconstruct the quark mass matrices.

The allowed regions of $f^{}_{\rm u}$ and $d^{}_{\rm u}$ are given in Fig.~1, where the $1\sigma$ ranges of $|V^{}_{ub}|$ and $|V^{}_{cb}|$ are also shown as the areas between the two horizontal dashed lines. In the upper panel, the gray points stand for the allowed values of $|V^{}_{ub}|$ and $f^{}_{\rm u}$, when the experimental constraint from $|V^{}_{us}|$ in Eq.~(\ref{eq:constraint}) is imposed. If all the four constraints are considered, we find that only the values of $f^{}_{\rm u} \in [0.16, 0.34]~{\rm GeV}$ survive. Using the quark mass ratios in Eq.~(\ref{eq:massratio}) and $m^{}_t = 172.1~{\rm GeV}$, one can immediately verify that $f^{}_{\rm u}$ is on the order of $\sqrt{m^{}_u m^{}_t} \approx 0.49~{\rm GeV}$. In the lower panel, $|V^{}_{cb}|$ and $d^{}_{\rm u}$ are shown, where one can observe that $|V^{}_{cb}|$ requires a nonzero value of $d^{}_{\rm u}$ in $[0.42, 2.16]~{\rm GeV}$. These results have confirmed our previous observations via approximate and analytical formulas. In addition, the allowed ranges of two phase parameters are $\phi^{}_1 \in [60^\circ, 105^\circ]$ and $\phi^{}_2 \in [0^\circ, 20^\circ] \cup [248^\circ, 360^\circ]$. The quark mass matrices compatible with experimental data at the $1\sigma$ level are reconstructed as follows
\begin{figure}[t]
\begin{center}
\includegraphics[scale=1.0]{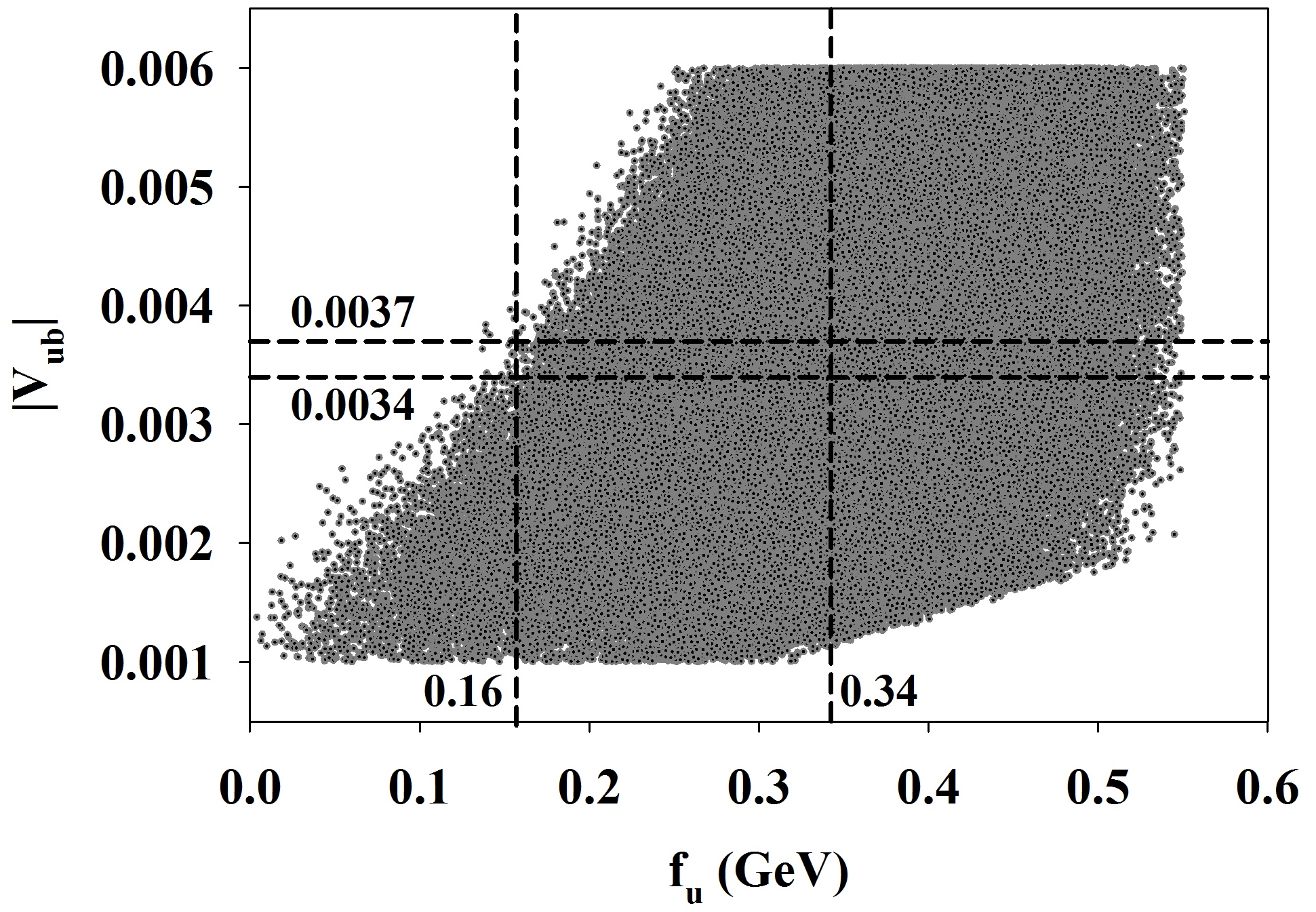}
\includegraphics[scale=1.0]{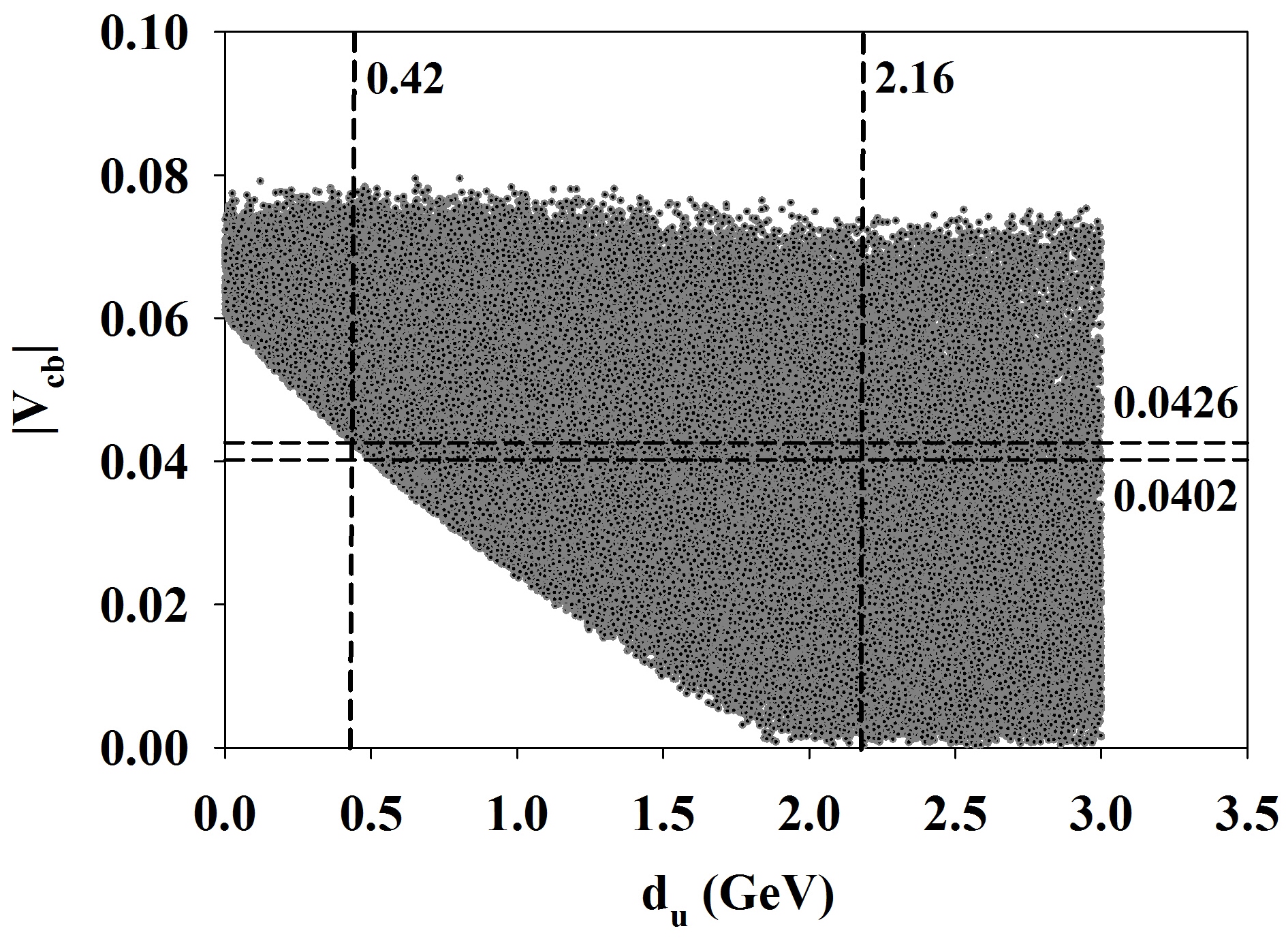}
\end{center}
\vspace{-0.4cm}
\caption{Numerical illustrations for the dependence of $|V^{}_{ub}|$ on $f^{}_{\rm u}$ (upper panel) and for that of $|V^{}_{cb}|$ on $d^{}_{\rm u}$ (lower panel).}
\end{figure}
\begin{figure}
\includegraphics[scale=1.0]{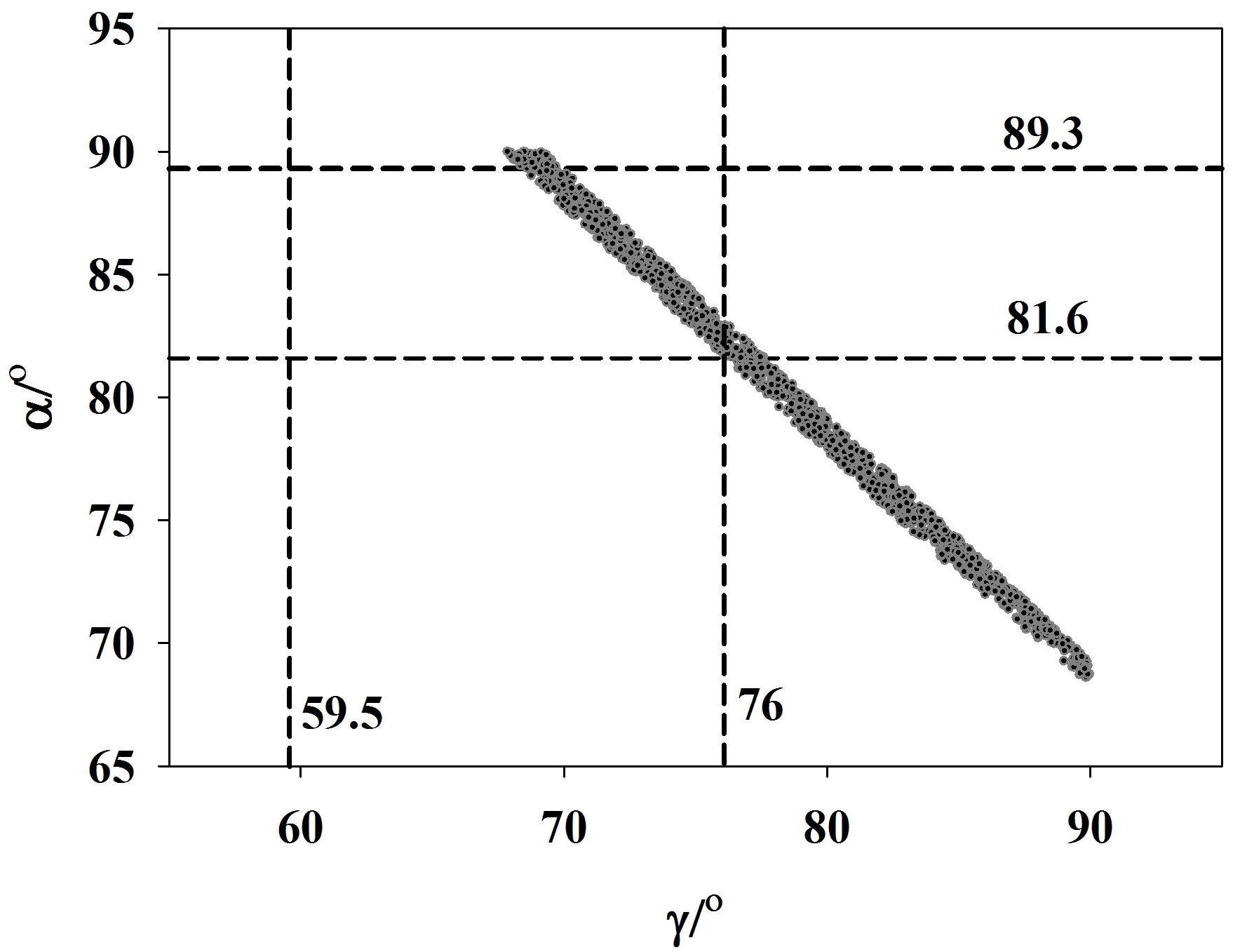}
\caption{The predictions for the two inner angles $\alpha$ and $\gamma$ of the unitarity triangle, where the experimental results in Eq.~(\ref{eq:constraint}) have been implemented to fix four model parameters.}
\end{figure}
\begin{eqnarray}
M^{\prime}_{\rm u} &=& \left(\matrix{{\bf 0} & {\bf 0}\cdots 0.02 & 0.2\cdots 0.3 \cr ~ & 0.4\cdots 2 & 14\cdots 22 \cr ~  & ~ & 171\cdots 173\cr}\right) {\rm GeV}\; , \nonumber \\
M^{\prime}_{\rm d} &=& \left(\matrix{{\bf 0} & 0.01\cdots 0.02 & {\bf 0} \cr
                                   ~ & {\bf 0} & 0.4\cdots 0.5 \cr
                                   ~ & ~ & 2.8\cdots 3.0\cr}\right) {\rm GeV} \; ,
\end{eqnarray}
where the off-diagonal elements of the real and symmetric matrices $M^{\prime}_{\rm u}$ and $M^{\prime}_{\rm d}$ have been omitted. A numerical investigation of the above mass matrices, without using the condition of factorizable phases, has been more recently carried out in \cite{Giraldo:2011ya, Giraldo:2015cpp} pointing to possible hierarchical structures for these matrices.

In Fig.~2, the predictions for the other two inner angles $\alpha$ and $\gamma$ are given, since the free parameters $\varepsilon^{}_{\rm u}$, $d^{}_{\rm u}$, $\phi^{}_1$ and $\phi^{}_2$ are restrictively constrained by the experimental data. Currently, the angle $\alpha = {85.4^\circ}^{+3.9^\circ}_{-3.8^\circ}$ has been extracted from the CP violation in the $B \to \pi \pi$, $\rho \pi$ and $\rho \rho$ modes, while $\gamma = {68^\circ}^{+8.0^\circ}_{-8.5^\circ}$ from that in the $B^\pm \to D K^\pm$ and $B^0 \to D^{*\pm} \pi^\mp$ decay modes. Thus, the precise determinations of $\alpha$ and $\gamma$ in $B$ factories and future collider experiments (e.g., LHCb) will test the unitarity of the CKM matrix. Additionally, we also find the Jarlskog invariant \cite{PhysRevLett.55.1039, Jarlskog:1985cw, Wu:1985ea}  $J^{}_{\rm CP} \equiv {\rm Im}\left[V^{}_{ts} V^*_{tb} V^*_{us} V^{}_{ub}\right] = (2.8\cdots 3.4)\times 10^{-5}$, which is in a perfect agreement with the data $J^{}_{\rm CP} = 3.06^{+0.21}_{-0.20}\times 10^{-5}$~\cite{Agashe:2014kda}. We also reconstruct below the viable quark mass matrices corresponding to an interchange of textures for the above case, and observe an agreement with hierarchical structures in Eq. (\ref{eq:hierarchical}), e.g.
\begin{eqnarray}
M^{\prime}_{\rm u} &=& \left(\matrix{{\bf 0} & 0.02\cdots 0.03 & {\bf 0} \cr                           ~ & {\bf 0} & 9.8\cdots 10.8 \cr ~ & ~ & 171\cdots 173\cr}\right) {\rm GeV} \;, \nonumber \\
M^{\prime}_{\rm d} &=& \left(\matrix{{\bf 0} & 0.01\cdots 0.02 & 0.003\cdots 0.01 \cr ~ & -0.07\cdots -0.04 & 0.07\cdots 0.21 \cr ~  & ~ & 2.8\cdots 3.0\cr}\right) {\rm GeV}\;,\nonumber \\
\end{eqnarray}
and the following allowed ranges for the two phases are obtained i.e. $\phi^{}_1 \in [70^\circ, 112^\circ]$ and $\phi^{}_2 \in [30^\circ, 50^\circ]$.

\subsection{Case II}
\begin{figure}
\includegraphics[scale=1.0]{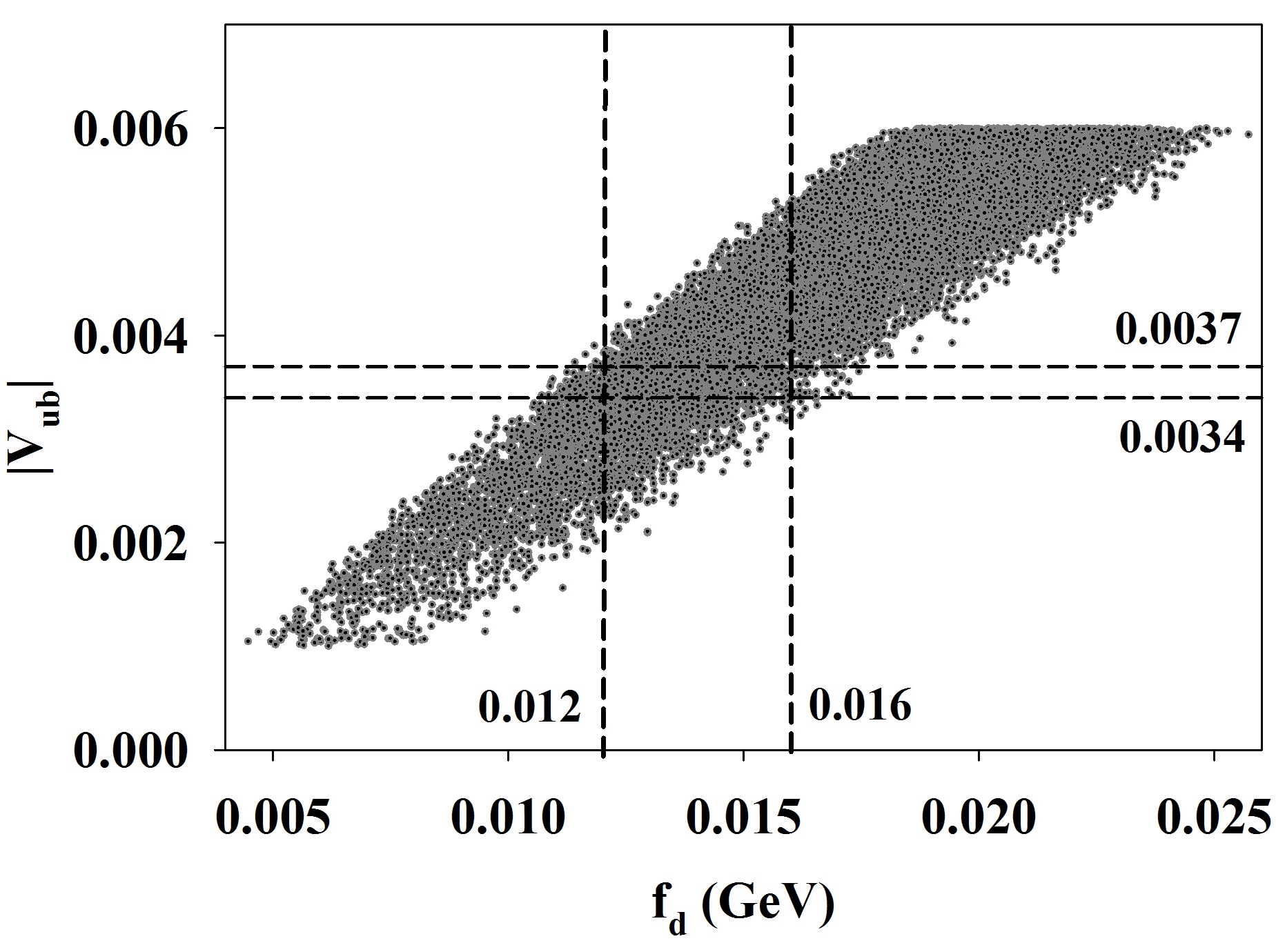}
\includegraphics[scale=1.0]{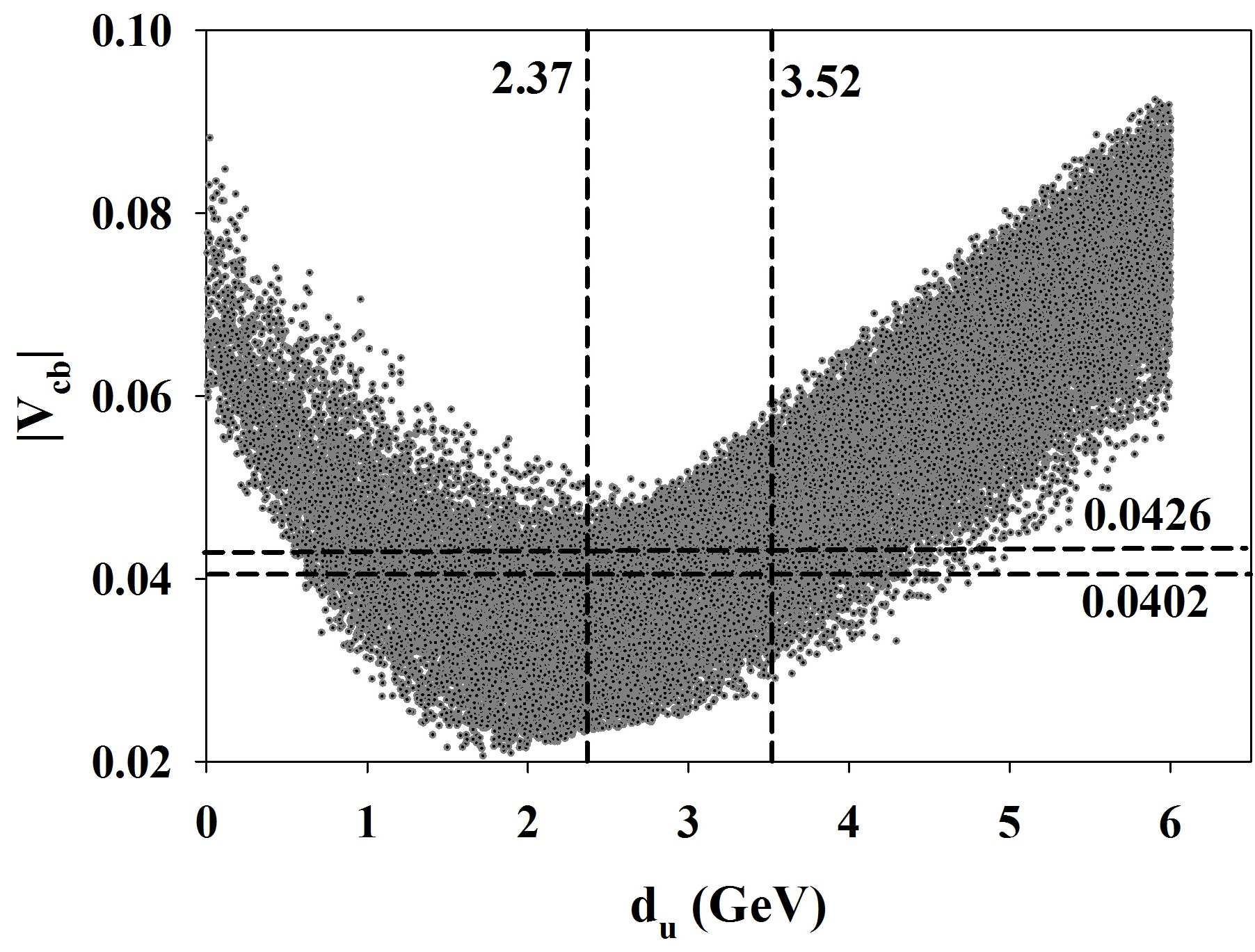}
\caption{Numerical illustrations for the dependence of $|V^{}_{ub}|$ on $f^{}_{\rm d}$ (upper panel) and for that of $|V^{}_{cb}|$ on $d^{}_{\rm u}$ (lower panel).}
\end{figure}
Now we turn to the second case in Eq.~(\ref{eq:caseII}) with $d^{}_{\rm u} \neq 0$, $\varepsilon^{}_{\rm u} = 0$, $d^{}_{\rm d} = 0$ and $\varepsilon^{}_{\rm d} \neq 0$.  Hence, there are also four free parameters $\{d^{}_{\rm u},\varepsilon^{}_{\rm d}\}$ and $\{\phi^{}_1, \phi^{}_2\}$. As in the previous case, these model parameters are allowed free variations within the corresponding allowed parameter spaces governed by Eqs.~(\ref{eq:hierarchical}) and (\ref{eq:approx}). With the help of experimental constraints in Eq.~(\ref{eq:constraint}), we obtain the following quark mass matrices that are consistent with the data at the $1\sigma$ level:
\begin{eqnarray}
M^{\prime}_{\rm u} &=& \left(\matrix{{\bf 0} & 0.02\cdots 0.03 & {\bf 0} \cr ~ & 2.4\cdots 3.5 & 23\cdots 26 \cr ~  & ~ & 168\cdots 171\cr}\right) {\rm GeV}\; , \nonumber \\
M^{\prime}_{\rm d} &=& \left(\matrix{{\bf 0} & 0.010\cdots 0.014 & 0.01 \cdots 0.02 \cr ~ & {\bf 0} & 0.3\cdots 0.4 \cr ~  & ~ & 2.8\cdots 3.0\cr}\right) {\rm GeV}\; .~~~~~
\end{eqnarray}

In addition, the allowed regions of two phases turn out to be $\phi^{}_1 \in [250^\circ, 300^\circ]$ and $\phi^{}_2 \in [13^\circ, 18^\circ]$. In order to understand the dependence of the CKM matrix elements on the nonzero elements $d^{}_{\rm u}$ and $f^{}_{\rm d}$, we show the values of $(|V^{}_{ub}|, f^{}_{\rm d})$ and $(|V^{}_{cb}|, d^{}_{\rm u})$ in the upper and lower panels of Fig.~3, respectively. Here we emphasize the importance of a non-vanishing $f^{}_{\rm d}$ for $|V^{}_{\rm ub}|$ and observe that $f^{}_{\rm d} \sim \sqrt{m^{}_d m^{}_s} \sim a^{}_{\rm d}$ accounts for the observed $|V^{}_{\rm ub}|$ within the $1\sigma$ interval. This is understandable from Eq.~(\ref{eq:approx}), since the down-type quark masses are less hierarchical than the corresponding masses of the up-type quarks. The allowed regions of the three quark flavor mixing angles and the three inner angles of the unitarity triangle remain the same as in the previous case. The Jarlskog invariant $J^{}_{\rm CP} = (2.82 \cdots 3.44)\times 10^{-5}$ is also well consistent with the experimental observations. For the purpose of completion, we present below the viable structures for the quark mass matrices when the texture zeros are interchanged in the above case, e.g.
\begin{eqnarray}
M^{\prime}_{\rm u} &=& \left(\matrix{{\bf 0} & 0.04\cdots 0.06 & 0.2 \cdots 0.4 \cr ~ & {\bf 0} & 9\cdots 11 \cr ~  & ~ & 171\cdots 173\cr}\right) {\rm GeV}\; , \nonumber \\
M^{\prime}_{\rm d} &=& \left(\matrix{{\bf 0} & 0.01\cdots 0.02 & {\bf 0} \cr ~ & -0.06\cdots -0.04 & 0.07\cdots 0.2 \cr ~  & ~ & 2.8\cdots 3.0\cr}\right) {\rm GeV}\;
 .~~~~~
\end{eqnarray}
along with $\phi^{}_1 \in [70^\circ, 106^\circ]$ and $\phi^{}_2 \in [35^\circ, 45^\circ] \cup [315^\circ, 330^\circ]$. Again, it is trivial to establish an agreement with the hierarchical structure of Eq.~(\ref{eq:hierarchical}).

\section{Conclusions}

In the present paper, based on the weak-basis transformations~\cite{Branco:1988iq}, we argue that the non-parallel four-zero textures of quark mass matrices are as general as the parallel ones, which have attracted a lot of attention in recent times \cite{Sharma:2015gfa}. Moreover, one advantage of the non-parallel textures is that these offer a natural translation of the observed strong hierarchies in the quark masses and mixing angles onto the corresponding mass matrix elements.  A phenomenological analysis of generic quark mass matrices with non-parallel textures incorporating possible extension of Fritzsch ansatz has been carried out under the assumption of factorizable phases. We are able to extract vital clues towards the formulation of these matrices using the available data on quark masses and CKM parameters in a model-independent way.  Two interesting observations are made. First, the experimental data from $|V_{ub}|$ measurements requires a nonzero element in the $(1, 3)$ position of quark mass matrices, namely, $f_{\rm u} \propto \sqrt{m_{\rm u} m_{\rm t}}$ or $f_{\rm d} \propto \sqrt{m_{\rm d} m_{\rm s}}\sim a_{\rm d}$, which is also crucial in realizing such hierarchical structures. Second, the experimental constraints from  $|V_{cb}|$ measurements forbid $d^{}_{\rm u}=d^{}_{\rm d} =0$ in these quark mass matrices.

We also derive general empirical relations for the CKM matrix elements, which offer a natural description of $|V^{}_{us}| \simeq |V^{}_{cd}|$, $|V^{}_{cb}| \simeq |V^{}_{\rm ts}|$ and $|V^{}_{ub}|/|V^{}_{cb}| < |V^{}_{td}|/|V^{}_{ts}|$. The  study indicates that there should exist, not one \cite{Sharma:2015gfa}, but several viable sets of hierarchical and Hermitian quark mass matrices, which may provide further possible clues for model building in top-down approaches \cite{Cheng:1987rs, Babu:1994kb, Desai:1997hk, Sarkar:2004ww, Carcamo:2006dp, DiazCruz:2009ek, BarradasGuevara:2010xs, HernandezSanchez:2013xj}. In this context, it has been recently shown \cite{Xing:2015sva} that the texture zeros in the quark sector are essentially stable against the one-loop renormalization-group-evolution of energy scale required in the top-down investigations of such mass matrices. It should be interesting to see if those specific structures of quark mass matrices can be verified in the future $B$-physics experiments.
\\
\begin{acknowledgments}
We are grateful to Prof. Z. Z. Xing for valuable discussions and suggestions. This work was supported in part by the Natural Science Foundation of China under grant No. 11375207 by the President's International Fellowship of CAS, by the
National Youth Thousand Talents Program, and by the CAS Center for
Excellence in Particle Physics (CCEPP).

\end{acknowledgments}
\bibliography{thebibliography}
\end{document}